\documentstyle[11pt,paspconf,epsf,psfig,twoside]{article}

\begin{document}

\title{The GOSSIP on the MCV V347 Pav.}

 \author{S.\,B.~Potter,$^{1}$ Mark Cropper$^{1}$ and P.\,J.~Hakala$^{2}$}
 \affil{$^{1}$Mullard Space Science Laboratory, University College London,
 Holmbury St. Mary, Dorking, Surrey, U.K.\\
 $^{2}$Observatory and Astrophysics Laboratory, FIN-00014, University of
 Helsinki, Finland. \\}

\begin{abstract}
Modelling of the polarized cyclotron emission from magnetic cataclysmic
variables (MCVs) has been a powerful technique for determining the structure of
the accretion zones on the white dwarf. Until now, this has been achieved by
constructing emission regions (for example arcs and spots) put in by hand, in
order to recover the polarized emission. These models were all inferred
indirectly from arguments based on polarization and X--ray light curves.

Potter, Hakala \& Cropper (1998)  presented a technique (Stokes imaging)  which
objectively  and  analytically  models the   polarized  emission to recover the
structure of   the cyclotron emission  region(s)  in MCVs. We  demonstrate this
technique   with the aid  of  a  test  case, then we    apply the technique  to
polarimetric  observations of   the AM Her   system V347  Pav.   As  the system
parameters  of V347 Pav  (for   example its inclination)   have  not been  well
determined, we describe an extension to the Stokes imaging technique which also
searches the system parameter space (GOSSIP).
\end{abstract}


\section{Introduction}
Many authors (e.g. Potter et al. 1997; Wickramasinghe \& Ferrario 1988;
Beuermann, Stella \& Patterson 1987; Cropper 1986) have found that the emission
region in MCVs must be extended into an arc or comma shape in order to explain
the shape of the polarized cyclotron light curves. Until now, the parameters
that define the shape and location of the emission regions have been adjusted
by a trial-and-error approach until the model gave a good fit. Clearly the
number of free parameters describing the location, shape and structure of the
emission regions becomes very large.

The technique of Potter, Hakala \& Cropper (1998, PHC) objectively modelled the
polarimetric data and  obtained  maps of the  emission  regions  for  the first
time. Their model had the  anisotropies of polarized emission incorporated into
it. This allowed them  to  model the  intensity, circular polarization,  linear
polarization  and position  angle data. Instead   of using the  maximum-entropy
method  and the conjugate optimization  (used  on intensity  data of Cropper \&
Horne   1994) they used  Tikhonov  regularization  and  a genetic algorithm for
optimisation.
 
\subsection{The cyclotron model}
The first step in modelling cyclotron emission from MCVs is to calculate the
viewing angle, i.e., the angle between the line of sight and the magnetic field
line from where the emission emanates. This was done by using a dipole magnetic
field formalism for the white dwarf, together with a system inclination and
offset dipole parameters (see Cropper 1989).  For a given magnetic longitude
and latitude of the emission point on the white dwarf, the viewing angle was
obtained for different phases of the spin cycle.  The local magnetic field was
calculated and subsequently the local optical depth parameter.  The intensity
spectrum and percentage of circular and linear polarization was then calculated
from interpolating on the cyclotron opacity calculations of Wickramasinghe and
Meggitt (1985).  Extended sources were modelled by summing the components of
many such emission points as a function of spin phase.

The optimisation of the model to the data proceeds by  adjusting the number and
distribution of emission points  across the surface  of the white dwarf. During
the optimisation  the  inclination, magnetic   dipole  offset (in latitude  and
azimuth) and the magnetic field strength at the poles are kept fixed.

\subsection{Stokes Imaging}

PHC used a genetic algorithm (GA) in order to optimise the fit to the
data. The GA works by first generating a set of random solutions. The
fitness of each solution is then calculated using

\begin{equation}
F(p)=\chi^2 + \lambda\sum_{i}{\|\nabla p_{i}\|^2}
\end{equation}

where $\|\nabla p_{i}\|$ is the mean gradient  of the number of emission points
at point {\it i} and $\chi^2$ is the chi-square fit.   The fitness is a measure
of how good the fit is to  the data plus the smoothness  of the image solution.
The solutions are ranked in  order of their  fitness.   The next generation  of
solutions are then  produced by a  type of natural  selection  procedure -- the
solutions that were ranked best are more probable  to breed the next generation
and the next population is generated by applying genetic crossover and mutation
operators to the selected parent pairs.

Eventually, after many generations, the improvement in fitness of the GA
solutions will start to level out and a more analytical approach is required to
improve the fit further. For the final approach to minimum the Powell's method
line minimisation routine (see Press et al. 1992) was used.

\subsection{The test case}
\begin{figure}[t]
\begin{center}
   \leavevmode
\psfig{file=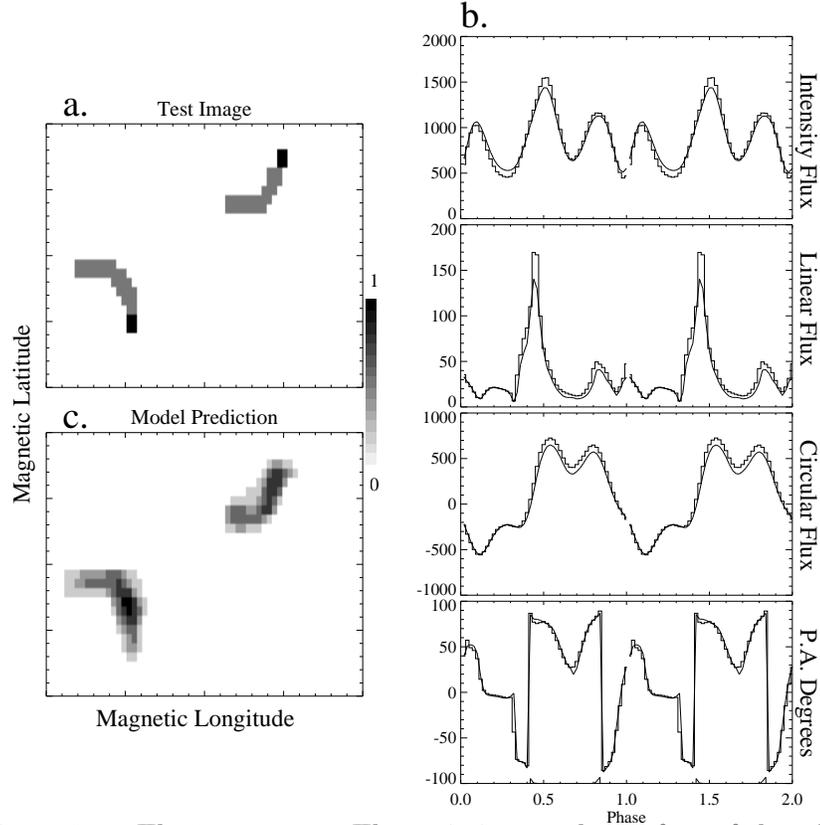,width=10cm,angle=0}
\caption{The test case {\bf a.} The emission on the surface of the white dwarf.
{\bf b.} The test emission light curves (histogram) and the model fit (solid
smooth curve). {\bf c.} The optimised image.}
\vspace*{-0.5cm} 
\end{center}
\end{figure}
To demonstrate   the   technique we have  generated   synthetic Stokes
parameter curves with 50    points  over the orbital   phase  (typical
datasets have  roughly this number). This gives  us 200 data points to
fit. In this paper we use a 6 degree resolution map on the white dwarf
surface, which in  turn yields  1740  free parameters ($60\times   29$
surface grid  elements  on  the white dwarf  surface).   The simulated
polarized  light curves presented   below were produced by calculating
the emission  that would arise from  two extended sources (Fig. 1a) in
the  B--band.  The   resulting light  curves  (Fig.  1b)  has infinite
signal--to--noise  ratio.    The  inclination  was  $80^{\rm o}$,  the
magnetic dipole  offset  was $10^{\rm o}$,  the  magnetic  polar field
strength  was $60$ MG and the  polar optical depth parameter $\Lambda$
was $1.0\times 10^{5}$.

Figure 1b shows the model fit (thick smooth curves) to the input test data
(histogram curves). The optimisation technique has reproduced the complex
features arising from the extended regions. These include the linear and
intensity features such as dips and peaks. It has also correctly modelled both
the positive and negative polarized light curves.

Figure 1c shows the model prediction for the shape and location of the
emission regions. The grey scale is a measure of the optical depth
across the emission region. As can be seen from the figure, the
optimisation routine has accurately located and mapped the emission
regions across the surface of the white dwarf.

\section{Genetically Optimised `Super' Stokes Imaging Procedure (GOSSIP)} 
For the test case described above we created polarimetric light curves
using   the cyclotron   model.     Therefore, the   system  parameters
(inclination, magnetic field orientation  and strength) were known. In
the case of real data these parameters are generally unknown.

PHC explained that the system parameters were  not included in the optimisation
procedure.  This is because these  parameters  are qualitatively different from
those describing  each    point on the   derived map.    Therefore,  the system
parameter space has to  be searched separately from   that searched during  the
`Stokes   imaging' technique, either  by trial  and error, or   by the use of a
genetic algorithm operating as an outer loop to the `Stokes imaging' procedure.
The outer  optimisation loop  optimises the final  fitness  values produced  by
several optimised  solutions.  In effect, the  optimised  solutions produced by
several  runs of the  `Stokes  imaging' technique become  solutions for genetic
optimisation themselves.

\subsection{The MCV V347 Pav}
\begin{figure}[t]
\begin{center}
   \leavevmode
\psfig{file=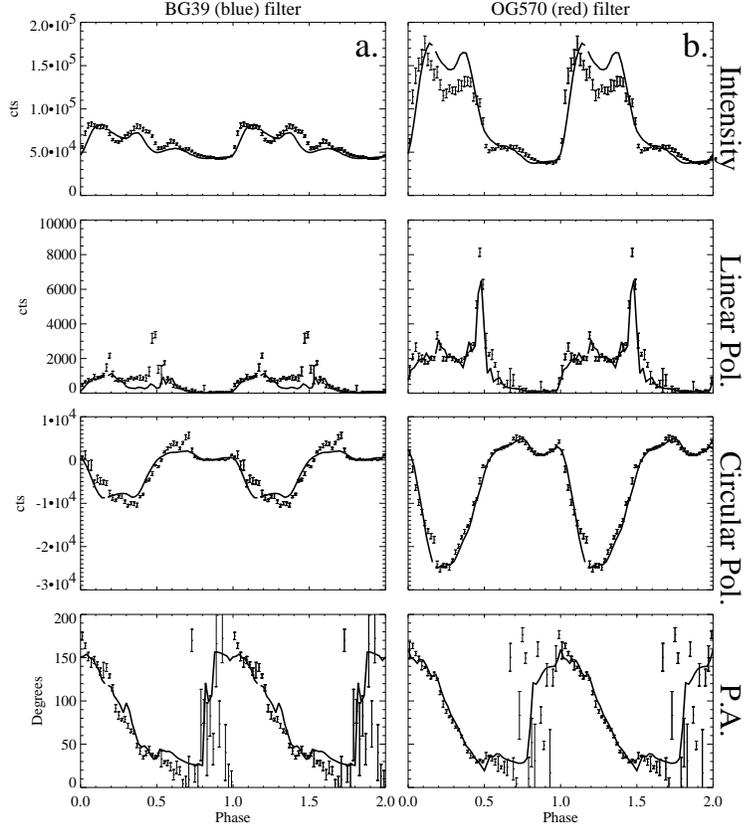,width=10cm,angle=0}
\caption{The `Stokes imaging' solution to the polarimetric observations of
Ramsay et al. (1996) for an inclination $=62^{\rm o}$ and a dipole offset
$=36^{\rm o}$.  {\bf a.} The blue data. {\bf b.} The red data.}
\end{center}
\end{figure}

Figure  2a,b shows the  blue and red  polarimetric observations of the
MCV  V347 Pav (RE J1844-741)  from Ramsay et  al (1996).  Bailey et al
(1995)  suggested that  the  positive and negative   excursions of the
circular polarization over  the orbital period of  $\sim $ 90 mins was
due  to emission arising from two  regions located at or near opposite
ends  of the magnetic  field. The high signal-to-noise observations of
Ramsay et al. (1996) make them ideal for the optimisation technique.

The technique predicts a best fit for an inclination of $62^{\rm o}$, dipole
offset of $32^{\rm o}$ and magnetic field strengths of 28 and 24MG for the
upper and lower poles respectively.  From Figure 2 it can be seen that the
optimised model solution has reproduced the red polarization data remarkably
well.  The morphology of all the red polarized light curves has been accurately
predicted, in particular, the variation in position angle, the relative amounts
of linear and circular flux, the peaks and dips in the circular polarization
and the overall variation in the intensity curve.  It has also predicted the
gross features of the blue polarized variations and the correct amount of
relative flux between the two wave bands.  The technique finds a solution that
fits both wave bands simultaneously.

However, it has not modelled the finer details of the blue polarimetric
variations.  This has arisen because of the less accurate determination for the
magnetic field strengths, thus affecting the quality of the fit to one
wave-band more than the other (the relative fluxes between the two wave-bands
depends on the magnetic field strength).  Also, the red polarimetric
observations have significantly more counts than the blue band. Therefore, the
optimisation routine will be more biased towards fitting the red data.  The fit
to the blue data can be improved by increasing the magnetic field strength on
the upper pole.  Maintaining a dipole field constraint, however, makes the fit
to the red data significantly worse.  Any temperature structure within the
shock will also affect the finer details of the polarized light curves in
different bands.  We have assumed a constant temperature of 10keV for the
emitting gas.

\begin{figure}
\begin{center}
   \leavevmode
\psfig{file=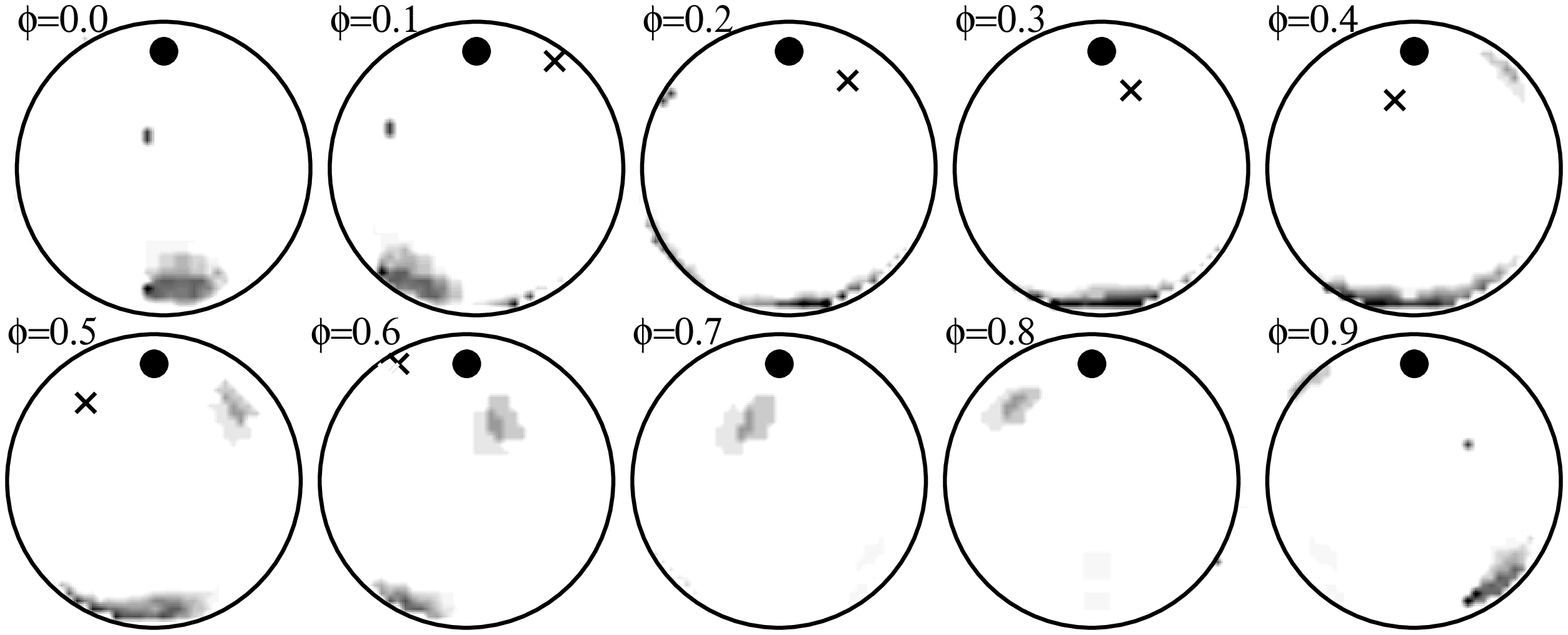,width=10.5cm,angle=-90}

  \caption{ The position of the cyclotron emission regions as viewed from Earth
for a complete orbital rotation. The dot and the cross represent the spin and
magnetic poles respectively.}
\end{center}
\end{figure}

Figure 3 shows the predicted emission regions mapped onto a sphere representing
the surface of the white dwarf as viewed from Earth for a complete orbital
rotation. The observed features of the polarimetric observations (Figure 2a,b)
can be explained with reference to Figure 3. The most prominent feature of the
observations is the bright and faint phase. The main emission region in the
lower hemisphere is responsible for the bright phase emission during $\phi \sim
0.0$--$0.5$ and the secondary smaller emission region is responsible for the
faint phase emission during $\phi \sim 0.5$--$1.0$. The south west--north east
orientation of the main emission region means that it gradually appears over
the limb of the white dwarf at $\phi \sim 0.0$ and then rapidly disappears at
$\phi \sim 0.5$ resulting in a large linearly polarized pulse. The secondary
emission region is visible for a larger fraction of the orbital period. It
first appears just before $\phi \sim 0.4$ and therefore contributes to the
large linearly polarized pulse. At $\phi \sim 0.8$ the magnetic field lines
feeding the secondary emission region are most pointing towards the line of
sight resulting in depolarization of the radiation by cyclotron
self-absorption. This is evident from the dip in the positive circular
polarization. The secondary region then moves away from the line of sight,
resulting in an increase in emission (seen as the contribution to the rapid
increase in the intensity) and the brief increase in positive circular
polarization at $\phi \sim 1.0$ . The secondary emission region continues to
stay on the upper limb of the white dwarf and gives rise to the linear pulse at
phase $\sim 0.2$ . The more gradual decline in the intensity from the bright
phase at $\phi \sim 0.5$ (seen especially in IR observations in Bailey et
al. 1995) can be attributed to the presence of the secondary region.

\subsection{The accretion region}
\begin{figure}[t]
\begin{center}
   \leavevmode
\psfig{file=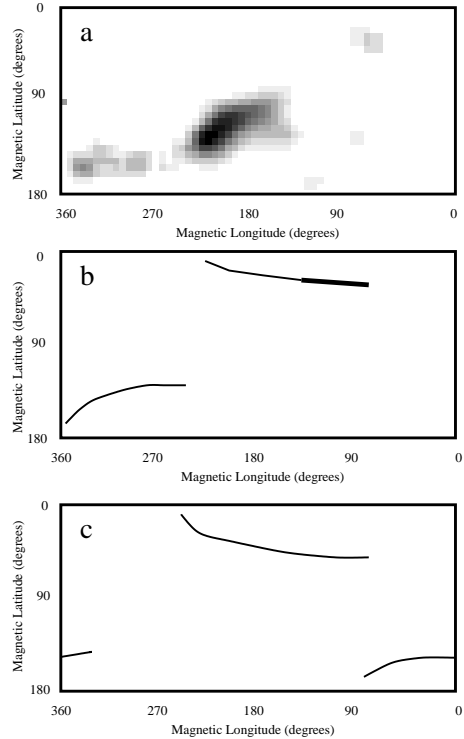,width=6cm,angle=0}

  \caption{A comparison between: {\bf a.} The optimised image solution. {\bf
b.} The model prediction in Ramsay et al. (1996) {\bf c.} The model prediction
of Bailey et al. (1995).}
\end{center}
\end{figure}

In Figure 4 we compare our prediction for the shape and location of the
cyclotron emission region with that of previous work. The figures are mapped
onto a flat surface. As a result, the large differences in longitude near the
polar regions are much smaller on the surface of a sphere than they appear on
the maps. The emission region is mapped onto the same coordinate system used in
each case. Figure 4b shows the prediction for the shape and location of the
emission region from Ramsay et al.  (1996).  The main difference is that Ramsay
et al.  (1996) located the main emission region responsible for the negative
polarization on the upper hemisphere of the white dwarf.  They also used a much
higher value for the inclination ($80^{\rm o}$) as an explanation for the equal
orbital coverage of the bright and faint phases and the variation in the
position angle.  Furthermore, they used two long thin arcs.  The only
similarity is that both models predict enhanced brightness on the trailing edge
of the main emission region.

Figure 4c shows the model prediction of Bailey et al. (1995). They also
predicted two long thin arcs located in approximately the same locations (large
changes in magnetic longitude near the poles, translate to small changes in
actual position) as that of Ramsay et al. (1996). However, they locate the main
emission region in the lower hemisphere in agreement with the `Stokes imaging'
technique. Furthermore, Bailey et al. (1995) estimated an orbital inclination
of $i=60^{\rm o}$ and a magnetic dipole offset of $\beta=50^{\rm o}$, also in
closer agreement with the `Stokes imaging' technique.

\section{Summary}
In this paper we have applied an extension of the Stokes imaging technique,
developed and discussed in PHC, simultaneously to the blue (3500--6500\AA \ )
and red (5500--9500\AA \ ) polarimetric observations of the AM Her star V347
Pav from Ramsay et al. (1996). We have also described an extension to the
Stokes imaging technique, which also searches the system parameter space.

Our technique has predicted an extended main cyclotron emission region, located
in the lower hemisphere of the white dwarf, and a secondary, smaller less dense
region diametrically opposite. The main emission region is found to consist of
a higher density region whose center is located $\sim 25^{\rm o}$ from the
magnetic equator.  It has a broken extended region extending ahead in phase
towards but trailing the magnetic pole.  The technique also predicts a
secondary emission region in the upper hemisphere which is relatively less
dense than the main emission region.  It is smaller in extent and almost
diametrically opposed to the center of the main emission region.  These maps
are different from those from previous less objective techniques in which two
long thin arcs were used to reproduce the polarimetric variations.

\acknowledgments SBP acknowledges  the PPARC for financial support  in
the form of a research studentship. PJH is supported  by an Academy of
Finland research fellowship.


\begin{references}
\reference Bailey, J. A. et al., 1995, \mnras, 272, 579
\reference Beuermann, K., Stella, L. \& Patterson, J., 1987 \apj, 316, 360
\reference Cropper, M. S., 1989, \mnras, 236, 935
\reference Cropper, M, \& Horne, K. 1994, \mnras, 267, 481
\reference Potter, S. B., Cropper, M, Mason, K. O., Hough, J. H. \& Bailey,
J. A., 1997, \mnras, 285, 82
\reference Potter, S. B., Hakala, P. J. \& Cropper, M, (PHC) 1998, \mnras 297,
1261
\reference Press, W. H., Teukolsky, S. A., Vetterling W.T. \& Flannery,
B. P., 1992, Numerical Recipes in FORTRAN, Cambridge University Press, p406
\reference Ramsay, G., Cropper, M., Wu, K. \& Potter, S., 1996, \mnras,
282, 726
\reference Wickramasinghe, D. T., \& Ferrario, L., 1988 \apj, 334, 412
\reference Wickramasinghe, D. T., \& Meggitt, S. M. A., 1985, \mnras, 214, 605

\end{references}
\end{document}